# Designing, Playing, and Performing with a Vision-based Mouth Interface

Michael J. Lyons
ATR Media Information Science Labs
2-2-2 Hikaridai, Seika-cho,
Soraku-gun, Kyoto, Japan 619-0288
mlyons@atr.co.jp

Michael Haehnel
Dept of EE and IT
RWTH Aachen University
Aachen, Germany

Nobuji Tetsutani
ATR M.I.S. Labs
2-2-2 Hikaridai, Seika-cho,
Soraku-gun, Kyoto, Japan

## ABSTRACT
The role of the face and mouth in speech production as well as non-verbal communication suggests the use of facial action to control musical sound. Here we document work on the Mouthesizer, a system which uses a headworn miniature camera and computer vision algorithm to extract shape parameters from the mouth opening and output these as MIDI control changes. We report our experience with various gesture-to-sound mappings and musical applications, and describe a live performance which used the Mouthesizer interface.

## Keywords
Video-based interface; mouth controller; alternative input devices.

## 1. INTRODUCTION
Articles on new interfaces for computer music often begin with a call for greater *embodiment* in the way computers are operated. This claim is sometimes backed up by citing developments in cognitive science which stress the importance of physical and physiological context for understanding the mind [25]. Similar considerations may be applied in the domain of machine-mediated human interaction. Current ways of interacting with computers neglect most of the physiology of human-human interaction and are surely unsuitable for most forms of communication, especially expressive forms such as music.

Working at McGill University half a century ago, Wilder Penfield and his colleagues [19] mapped the sensory-motor cortex by electrical stimulation of conscious patients undergoing neurosurgery. Their pictorial summary of the findings, the somatosensory and motor homunculi, are widely known and their importance for human-machine interaction [2] as well as musical interfaces [8] has been recognized. A striking aspect of the motor homunculus (see Figure 1) is the relatively large area devoted to the organs critical for verbal and non-verbal human communication: the lips, mouth, tongue, larynx, and the face.

The importance of the face and especially the mouth in communication inspired us to develop a musical controller which takes input from facial actions. The face, especially the mouth area, is involved both in sound production, in speech, singing, and in non-verbal emotional communication, in facial expression. It therefore seemed interesting to attempt to design a musical interface making use of our expertise for muscular action of the face.

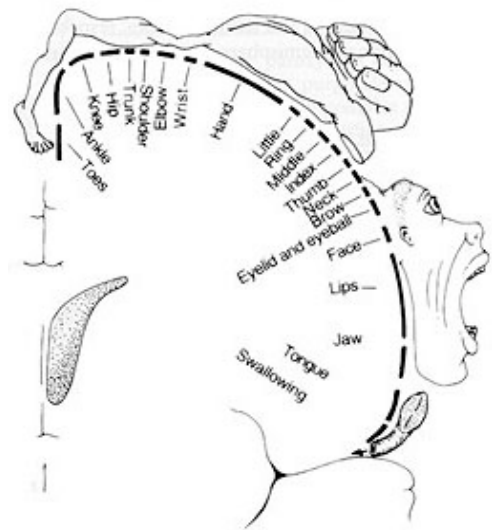

**Figure 1. The motor homunculus or representation of body areas in the motor cortex (adapted from [19]).**

This paper reports work using a video-based approach and focuses on the area of the mouth. Preliminary results of the study were previously published in brief format [13]. The current article is intended as a more complete record of the project in which we: (a) state the context of the work by reviewing related studies (section 2); (b) describe the implementation in detail including more recent developments, discussing design considerations as well as lessons learned (section 3); (c) report our experience with several mappings and musical applications of the controller (section 4); and (e) describe a public performance in which the controller was used (section 5); and (f) conclude with general observations (section 6).

## 2. RELATED WORK
### 2.1 Mouth & Vocal Tract Controllers
The fact that oral cavity shape influences the human voice means there are complex neural circuits relating for muscular control the mapping of shape to sound effect. Use of the oral cavity for modulating sounds other than those produced by the vocal tract is probably as old as instrumental music itself, evidenced by the presence of instruments like the mouthbow in folk cultures around the world.

Functioning by the same principle as acoustic mouth controllers, the TalkBox, which enjoyed popularity in the 1970's, allows a player to directly filter an audio signal with the acoustic transfer of the oral cavity. Holding a small speaker in the mouth, the player modulates the signal by varying the oral cavity shape and position of the tongue. Since the actual acoustic properties of the mouth modify the signal, the TalkBox is intuitive to use. However the range of sound effects is limited by the same acoustic possibilities. It is also requires the player/singer to keep the device in their mouth.

The Vocoder [5] allows effective vocal tract control of synthesized electronic sounds via audio signal processing extraction of voice parameters to modulate synthesized sounds. Using a Vocoder is more akin to speaking or singing than to playing an instrument since it is activated by sounds produced in the vocal tract itself.

By contrast, the interface developed by Orio [16] probes the shape of the oral cavity by stimulation with an external acoustic source. Shape parameters extracted from analysis of the response are output as MIDI controls. Orio found that users could easily learn to control two independent parameters by varying mouth shape, but greater difficulty controlling three parameters.

Vogt *et al.* [26] used ultrasound imaging to measure tongue position and movement in real-time for sound synthesis control. With the Tongue 'n' Groove, an ultrasound device is held below the jaw and an image of the tongue contour reconstructed, or alternatively, optical flow due to tongue motion is calculated. Several mapping metaphors were explored; *e.g.* tongue position was used to play a physical model of the singing voice.

## 2.2 Vision-Based Musical Interfaces

Several previous works have used computer vision techniques for musical interaction.

Multimedia installation artist David Rokeby has experimented for many years with his Very Nervous System [22], or VNS, which is now available for purchase. The VNS web pages do not give an explicit description of what it computes, but VNS appears to be based primarily on pixel calculations responsive to movement, such as temporal differencing in user defined trigger zones.

The BigEye software [23], available commercially from STEIM, allows tracking of coloured objects against a set of definable regions in the video frame, with variables such as relative position, size, and speed output as MIDI parameters.

The EyesWeb software platform [1,6], freely available from the InfoMus group at the University of Genoa, includes several computer vision modules allowing tracking of objects and coloured blobs as well as modules for estimation of affective and expressive qualities of movements, with several output options including MIDI and OSC.

Some vision-based controllers adapt software developed for non-musical purposes. The DanceSpace system [18] added to the MIT Media Lab's Pfinder vision-based person tracker, to allow mapping of movements of tracked limbs, torso, and to changes in musical parameters.

The Augmented Groove system [20] uses the University of Washington HIT Lab's Augmented Reality (AR) Toolkit which supports tracking of high-contrast two dimensional patterns. The AR Toolkit can extract translation, rotation, and tilt of objects labeled with the patterns. The Augmented Groove system maps tracking parameters to MIDI control changes, allowing users to modulate sequencer loops by manipulating physical objects.

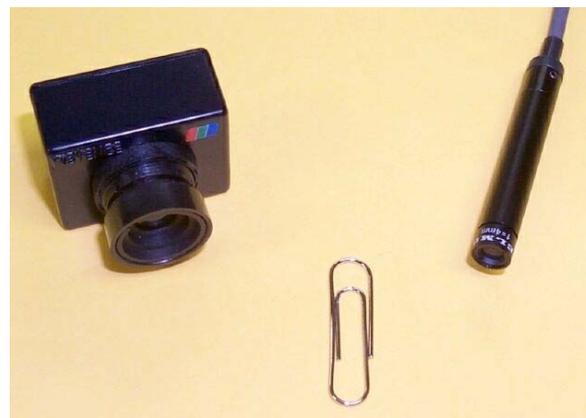

**Figure 2. Two miniature CCD cameras used in this work.**

Those with the resources often prefer to develop specific software for a project as this allows greater control over how the solution is implemented.

The Iamascope's vision-to-music subsystem [7] divides the video input frame into detection zones mapped to notes of a chord. A motion detection algorithm allows note triggering with free gestures. The kaleidoscope subsystem mirrors the video input to display intriguing visual feedback of the user's own gestures.

With the Imaginary Piano [24], video input from a camera facing the player is analyzed with a motion detection algorithm which responds to movement of the hands below a vertical threshold to trigger piano notes having pitch determined by the horizontal coordinate.

Ng [15] seems to be the only other work to suggest using a vision-based interface to transform facial gestures to MIDI controls. A video demo of their prototype was shown at NIME-02, but details of their implementation have not yet been published.

## 3. DESIGN EVOLUTION
## 3.1 Face Tracking System

At the outset, our aim was to utilize actions from several areas of the face for musical expression, including movements of the eye regions, eyebrows, mouth, cheeks and movements of the whole head. We started by building upon a vision-based face tracking system developed in our group at ATR. The first prototype was implemented on an SGI O2 computer, using the O2's built-in framegrabber and the IRIS video library. This allowed acquisition of NTSC quarter-frame images (320x240 pixels) at 15 fps. This is the minimum useable frame rate for most musical applications: latencies are noticeable but still tolerable.

We initially considered using a more sophisticated feature shape representation [11,17], but experiments showed that the shadow area of the mouth could be extracted by a very simple colour and intensity thresholding algorithm. First, a region large enough to include the mouth area with certainty is chosen, based on the inter-ocular distance from the face tracking module. Next, pixels in this region satisfying the following equation:

$$I < I_{min} \quad \text{and} \quad R > R_{max}$$

are segmented, where $I$ is pixel intensity and $R$ is its red component and $I_{min}$ and $R_{max}$ are set thresholds. With appropriate values for the thresholds, under a large region of lighting conditions most of the pixels satisfying this condition belong to the shadow area of the mouth. Segmentation by colour thresholding is widely used to track objects in vision systems. For example skin colour is used as a cue in many face detection algorithms, though it is widely known to be affected by the intensity and colour of the illumination. Thresholding of the mouth shadow area seems to be considerably more stable to such illumination changes because we are detecting the *absence of a surface*: the appearance of the cavity is more robust than that of the surrounding skin areas. Use of the automatic gain and colour balance control on the cameras adds to the robustness of the system to lighting changes.

The system has some limitations, for example, dark facial hair near the mouth may also be selected by the thresholding operation. With very dark skin, thresholds may need to be changed or additional lighting used.

The pixels obtained by the thresholding operation were analyzed using a principal components analysis [4] to find the major and minor axis of the segmented area, which approximates an elliptical blob. Use of an algorithm based on colour segmentation and invariant statistics of the pixel coordinates has the advantage that it is robust to translation and in plane rotation of the mouth region. Hence, movements of the camera, unavoidable due to slight vibration of the beam, do not strongly affect the shape parameters extracted from the image. Higher level algorithms based on tracking loops for position estimation, would not share this property.

The two shape parameters were mapped to two MIDI control change values. These were used to control timbre of various physically modelled instruments using the demos Perry Cook's Synthesis ToolKit [3] running on the same SGI O2. Segmented pixels were displayed in red on a video output of the players face. Experiments with this system quickly convinced us that the mouth functions well as a controller of synthesized sound and we next concentrated on developing a mouth controller for use in actual musical performance situations.

### 3.2 Headworn System

When we began this research, the face tracking module was limited to a video processing rate of 13 fps, and tracking was interrupted by large speed or amplitude of head movements. Tracking performance has since been increased to full frame rate, but robustness it is still not adequate for live performance situations. In addition, the apparent facial expression in 2D projection depends on head orientation [12]. Finally, experiments with the tracking system convinced us that a wearable system would allow performers greater mobility and comfort.

These considerations led us to concentrate research on a system based on a head mounted camera pointed directly at the mouth area (see Figure 3). Camera distance and focal length of the lens were chosen so that that the input video frames contained the facial region of the mouth, and excluded other areas that are picked up by thresholding such as the nostrils and, occasionally, a shadow below the lower lip. This eliminates the need for a face tracking system.

The headset is a modified Shure SM10A with the microphone and beam assembly removed and replaced with a miniature video camera mounted on a homebuilt aluminum arm. It is important to counterbalance the weight of the camera. Miniature, lightweight video cameras are now widely available (see Figure 2). Most of the work reported here used a Keyence CK-200B miniature colour CCD camera with standard NTSC analog output. We also tried the expensive Elmo QN42H camera pictured in the figure, which gave similar results. The Keyence camera is economical and ideally suitable for use with the Mouthesizer. A web site with more information on obtaining the camera is listed with the references [10].

### 3.3 Machine-Vision Board

To demonstrate the feasibility of an inexpensive, portable, and stable system we next implemented a hardware prototype. We selected the Cognachrome 2000, made by Newton Labs (Renton, WA), a dedicated machine vision board based on the Motorola 68322 processor. The Cognachrome tracks the position, size, aspect ratio, and orientation of several colour blobs at 60 Hz and a spatial resolution of 200x250 pixels, with a proprietary algorithm that uses colour and intensity thresholding and connected region analysis. Video input and output is NTSC format and data communication is via a serial port. Mouth shadow blob dimensions as detected by the board were remapped to two MIDI control changes via a program running on a desktop program. The Cognachrome is programmable allowing onboard implementation of MIDI communications. However, it has several disadvantages, the most important of which were that mouth shadow tracking was more sensitive to illumination changes than with the software algorithm we implemented and that the tracking algorithm could not be easily modified.

### 3.4 Current Implementation

Flexibility considerations led us to return to a software implementation, using Visual C++ and Direct-X running under Windows. The current Mouthesizer operates as a Direct-X filter, allowing use of the system with any input video device for which a driver is available.

Several improvements to the algorithm were made. Connected region analysis is applied after the colour and intensity thresholding operation. This removes thresholded pixels outside of the mouth shadow area. Two types of simple temporal filters were added to remove noise due to rapid fluctuations in lighting or shadows, illustrated in figure 4.

Filter A discards any output that differs from the average of the two most recent output values by more than a set threshold.

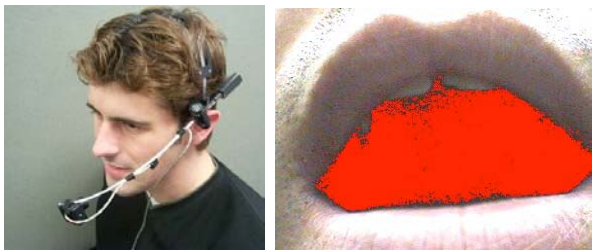

**Figure 3. The headset and a view from the camera.**

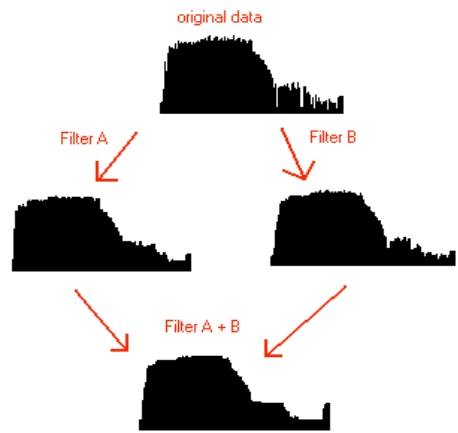

**Figure 4. Filters used with the Mouthesizer to remove noise.**

Filter B temporally smoothes the data by averaging current outputs with previous ones. The filters can be independently turned on or off while the software is running.

A desktop system on a Pentium II with a Winnov Videum capture card ran at 30 fps, while on a Pentium III notebook with I-O Data PCCAP video capture card the system ran at 15 fps, both at a resolution of 320x240 pixels. The system now also works with Firewire and USB cameras at full frame rate. Some recent palmtops should now have sufficient processing power to run the Mouthesizer algorithm at low resolution, which would allow it to be used as a fully wearable device.

## 4. MAPPING & APPLICATIONS

Below we first report gesture-to-sound mappings which were found to work well with the Mouthesizer. Then we describe actual applications of the Mouthesizer to playing music. Most of our experiments with mappings and applications were made with the Nord Modular Virtual Analog Synthesizer (from Clavia, Sweden). With the Nord Modular, synthesis or audio effects patches are edited in software with an intuitive graphical interface, but run on dedicated DSP hardware. Patch variables were easily adjusted using control panel knobs and driven by external MIDI controllers.

### 4.1 Mapping

In all cases the mouth shape parameters were mapped to two MIDI control changes. The mappings discussed below were not intended to be one-to-one mappings of shape to sound. Rather, two principles guided our experimentation: the role of the mouth as a filter in sound production and the action of the facial muscles in the facial expression of emotion. Our goal was to try to create intuitive and compelling mappings from action to sound by making use of existing motor expertise and brain maps for sound production and emotional expression.

Musical interface mapping is a subtle issue [9] and there is room for further exploration of the expressive potential of the Mouthesizer. For example, for some vocal consonants the lips and tongue act as sources of sound. Expression of certain emotions such as surprise or mirth can have relatively rapid onset dynamics, which may not be well modelled as continuously changing controls. Cursory experiments suggested that it should also be interesting to use the Mouthesizer interface to trigger sound events such as samples, but we have so far not pursued this line as a mapping strategy. To encourage further experimentation with mappings, we are planning to make a version of the Mouthesizer available in the near future.

#### 4.1.1 Mouth Height

One of the most is compelling and intuitive mappings we found uses the height of the mouth opening to control the cut-off frequency of a sweeping resonant low-pass filter. With this mapping opening the mouth opens up the filter, letting higher frequency components of the sound pass. This audio effect is popularly known as *wah-wah*, an onomatopoeic term describing the effect of opening and closing the mouth while voicing the sound "ah". This mapping mimics effects available with mouthbow and jaw harp instruments, as well as the TalkBox. Simple, intuitive effects also result from mapping mouth height to volume control, sustain, or damping.

#### 4.1.2 Mouth Width

We found interesting expressive effects by mapping mouth width to *distortion* level of an amplifier. This was motivated by the action of the mouth in expressions of pain, suffering, or fear. Opening the mouth increases the non-linearity of the response of an audio-amplifier which clips the guitar signal waveform. Stretching the corners of the mouth apart in a grimace increases the level of distortion.

We also tried mapping mouth width to the resonance of a resonant low-pass filter. Stretching the mouth wide gives a high-frequency chirp, expressing arousal without the negative emotional valence of the distortion effect.

#### 4.1.3 Mouth Aspect Ratio

Here the mouth aspect ratio or eccentricity was used to control an audio morph between formant filters for three of the fundamental vowel sounds [a], [i], [o]. [i] has the greatest eccentricity, [o] is the most rounded, having the least eccentricity, [a] has intermediate eccentricity. An existing formant filter module of the Nord was used to control the morph. This gives an intuitively natural mapping of mouth shape to filter audio effect which is similar to acoustic effects playable with controllers like the TalkBox.

### 4.2 Applications

The Mouthesizer was played informally, with three main musical applications, guitar effects, keyboard, and sequenced loops.

#### 4.2.1 Guitar Effects

Ichiro Umata, jazz guitarist and cognitive science researcher, used the Mouthesizer to control guitar effects. Mouth height controlled wah-wah as described above and mouth width adjusted the amount of distortion. This experiment used an early version of the Mouthesizer running at 15 fps on an SGI O2 computer. The guitarist had little prior practice using the Mouthesizer. After a session lasting approximately one hour, he noted that the

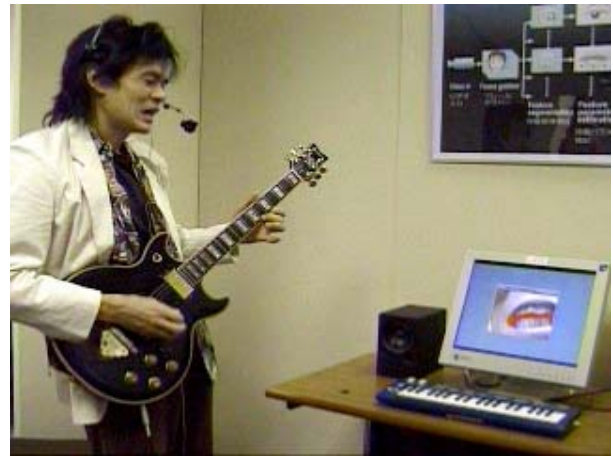

**Figure 5. Controlling guitar effects with the Mouthesizer.**

Mouthesizer was easily learned and more natural to play than a pedal controller. He also observed that changes of mouth width and height are correlated for most movements, making the controller more interesting to use than if the two audio effects were independently adjustable. This agrees with the findings of Hunt *et al.* in their study of simple and complex mappings.

#### 4.2.2 Keyboard Synthesizer Demo

We experimented with several keyboard synthesizer patches running on the Nord Modular, controlling patch parameters with the Mouthesizer. We tried these at a live demo during the ATR Open House exhibition. The mappings which easiest for most visitors to understand were simple ones usually associated with keyboard pedals such as volume control, sustain, and damping.

Reactions to the Mouthesizer varied greatly. Some visitors, having conservative musical tastes, found the concept strange or at least humourous. Others, sympathetic to electronic music, found it more appealing.

### 4.2.3 Sequenced Loops

Inspired by the Augmented Groove system, we used the Mouthesizer to control techno loops. This allows one to add expression to an automatically played musical sequence. Again, sweeping filters, resonance, distortion, and formant filter morphs work well here.

## 5. LIVE PERFORMANCE

Jordan Wynnychuk gave a 30 minute solo live performance of improvised electroacoustic music using the Mouthesizer at the Kyoto Kyoryukan to audience of about 30 people. Figure 6, a picture of a rehearsal for the performance shows the instrumentation used in the concert. Jordan used a touch-sensitive MIDI control pad and STEIM's LiSa (Live Samples) software, to trigger and manipulate samples with the fingers of both hands. The sounds used in this performance consisted of glitches, squeaks, blips, bangs, buzzes, whirs, other "error" sounds, as well as samples of percussion instruments.

Audio effects running on the Nord Modular were controlled using the Mouthesizer. Aesthetic considerations led us to experiment with audio effect mappings less intuitive than the ones described above, including a mix of extreme distortion, high and low pass filtering, and panning between left and right channels. One of the most interesting effects we discovered panned high and low frequency filtered versions of the audio signal between the two channels by opening and closing the mouth.

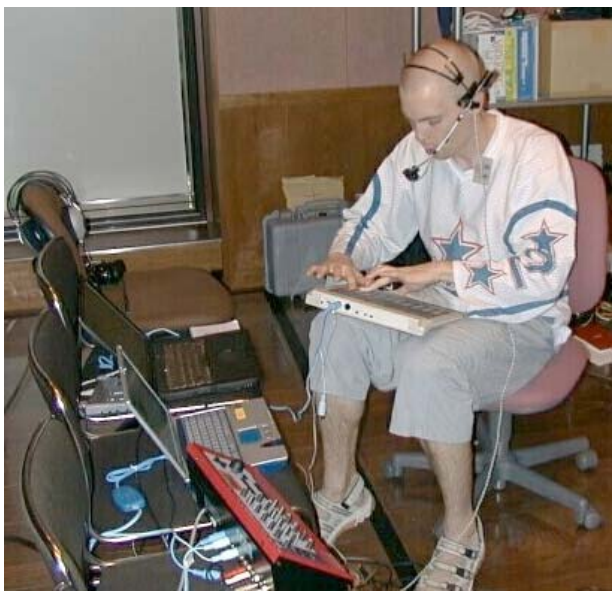

**Figure 6. Jordan Wynnychuk with controllers used for the live performance.**

Figure 7 shows a sequence of pictures from Jordan's performance in which a mouth gesture is being used to adjust sound expressively. Not visible in these images are the highlighted mouth shadow regions, which were projected on a screen beside the stage.

In addition to the interest and originality of Jordan's performance, audience members' attention seemed to be captured by the novelty of the Mouthesizer interface and concept. Many asked afterwards about how it worked or suggested ways it could be used to play music. One or two reported on their check for causality between mouth action and aural effect: they found it sometimes easily visible but quite obscure at other times. This appeared to be mainly a function of the mapping.

## 6. CONCLUSION

Experimentation with a variety of mappings and musical applications, and use of the Mouthesizer in a live public performance confirmed our hypothesis that facial gestures, especially movements of the mouth, are suitable for expressive musical control. Several lessons were learned in the design process.

Abandoning the head-tracking system early in the project later led us to avoid discrete classification of facial expressions for control of musical effects. This was fortunate: a controller which categorized emotion discretely would have limited responsiveness to movement quality and would reduce expressiveness. Rather, we captured a signal responsive to the motion of the mouth with a simple vision algorithm and relied on carefully chosen mappings to enable expressive control of sound production.

A valuable feature of the Mouthesizer is the visual feedback provided by highlighting segmented pixels. This gives a highly visible display mirroring the player's actions. Rizzolatti and colleagues [21] have discovered "mirror neurons" in monkey frontal lobes which respond to specific motor actions both when they are performed and when the same or a similar action is observed. Such circuits are thought to be important for learning motor behaviours. Controllers which mirror a motor action via visual display (and perhaps also auditory display?) should strongly stimulate these circuits. This may lend appeal to musical interfaces that mirror the player's gestures. A previous vision-based interface which shares this property is the Iamascope [7].

Additionally, visual perception of lip movement is known to affect auditory perception of speech [14] (the "McGurk effect"). The Mouthesizer brings such cross-modal sensory mappings into play for both performer and observer.

Such considerations lead us to conclude that video-based and other new gestural controllers offer much more than a visually engaging spectacle for the audience. They enable news ways to use the body and its sensory-motor systems to explore human expression and communication via sound.

Recently the Mouthesizer was integrated with an improved face tracking system to create an interface which allows users to point and click with facial gestures. Hence, work on a musical interface led to a non-musical side product. Musical applications seem to stimulate exploration of a very wide range of interaction paradigms. In this way, the NIME conference may have an important contribution to make to the wider field of human-computer interaction.

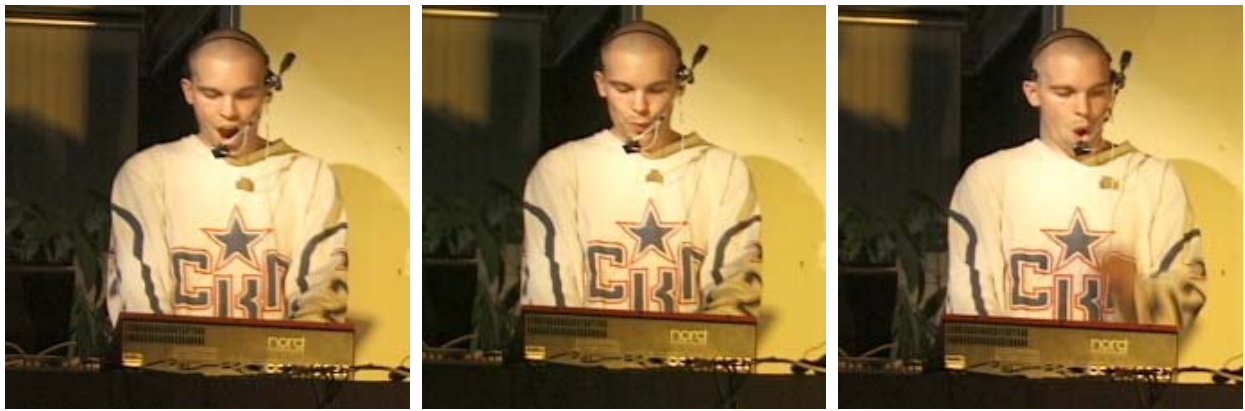

**Figure 7. Jordan Wynnychuk using the Mouthesizer in a live performance.**

## 7. ACKNOWLEDGMENTS

MJL thanks Christoph Bartneck, Rodney Berry, Palle Dahlstadt, Sidney Fels, Steven Jones, Ivan Poupyrev, Ichiro Umata, Jordan Wynnychuk, and Tomoko Yonezawa for stimulating and helpful interactions. This work was supported in part by the Telecommunications Advancement Organization (TAO) of Japan.